\documentclass[%
 twocolumn,
 reprint,
showpacs,
 amsmath,amssymb,
 aps,
]{revtex4}

\usepackage{graphicx}
\usepackage{dcolumn}
\usepackage{bm}

\begin{document}

\preprint{XXXX}

\title{Thermal Phase Transition of Generalized Heisenberg Models for SU(N) Spins on Square and Honeycomb Lattices}

\author{
Takafumi Suzuki${}^1$, Kenji Harada${}^2$, Haruhiko Matsuo${}^3$, Synge Todo${}^4$, and Naoki Kawashima${}^5$
}

\affiliation{
${}^1$Graduate School of Engineering, University of Hyogo, Hyogo, Himeji 670-2280, Japan\\
${}^2$Graduate School of Informatics, Kyoto University, Kyoto 615-8063, Japan\\
${}^3$Research Organization for Information Science and Technology, Kobe 650-0047, Japan\\
${}^4$Department of Physics, University of Tokyo, Tokyo 113-0033, Japan\\
${}^5$Institute for Solid State Physics, University of Tokyo,  Kashiwa 277-8581, Japan
}


%
\date{\today}

\begin{abstract}
We investigate thermal phase transitions to a valence-bond solid phase in SU(N) Heisenberg models with four- or six-body interactions on a square or honeycomb lattice, respectively.
In both cases, a thermal phase transition occurs that is accompanied by rotational symmetry breaking of the lattice.
We perform quantum Monte Carlo calculations in order to clarify the critical properties of the models. 
The estimated critical exponents indicate that the universality classes of the square- and honeycomb-lattice cases are identical to those of the classical $XY$ model with a $Z_4$ symmetry-breaking field and the 3-state Potts model, respectively. 
In the square-lattice case, the thermal exponent, $\nu$, monotonically increases as the system  approaches the quantum critical point, while the values of the critical exponents, $\eta$ and $\gamma/\nu$, remain constant.
From a finite-size scaling analysis, we find that the system exhibits weak universality, because the $Z_4$ symmetry-breaking field is always marginal. 
In contrast,  $\nu$ in the honeycomb-lattice case exhibits a constant value, even in the vicinity of the quantum critical point,  because the $Z_3$ field remains relevant in the SU(3) and SU(4) cases.
\end{abstract}

\maketitle


\section{Introduction}
\label{sec:Introduction}
The classification of various continuous phase transitions has been successfully discussed 
from the viewpoint of the Landau-Ginzburg-Wilson (LGW) paradigm~\cite{GL,GLW}.
The essential principles of the paradigm are the clarification of (local) order parameters and the characterization of breaking symmetries.
Recently, the possibility of deconfined critical phenomena (DCP)~\cite{SenthilVBSF2004,SenthilBSVF2004,SenthilBSVF2005} 
has attracted considerable attention as a quantum phase transition (QPT) beyond the LGW paradigm.
DCP have been predicted to occur at the QPT point between a magnetically ordered phase, such as the N\'eel phase, and 
the valence-bond solid (VBS) phase in two dimensional (2D) systems.
Remarkably, this phase transition is continuous, although the symmetry group in one phase is not the subset of another phase.
The well-known models that are expected to exhibit DCP are the generalized Heisenberg models with multibody interactions for SU(N) spins namely, SU(N) $JQ_m$ models~\cite{Sandvik2007}.
Considerable effort has been expended to numerically determine whether the QPT of this model family is of the second order or weak first order,
however, a satisfactory result has not yet been obtained~\cite{Sandvik2007,LouSK2007,MelkoRG2008,Sandvik2010,KaulS2012,KuklovAB2008,ChenK2013,HaradaK2013,PujariKDFA2013}.

An interesting aspect of DCP is that the transition may occur independently of the lattice geometry~\cite{SenthilBSVF2004,SenthilBSVF2005}.
In a previous study~\cite{HaradaK2013}, we evaluated the critical exponent, $\nu_{\rm QPT}$, at the QPT point between the N\'eel and VBS phase
in SU(N) $JQ_m$ models on both square and honeycomb lattices using quantum Monte Carlo (QMC) calculations.
From the finite-size scaling (FSS) analysis, we confirmed that $\nu_{\rm QPT}$ is independent of the lattice geometry but depends on the SU(N) symmetry.
This result strongly suggested the presence of DCP in the SU(N) $JQ_m$ models.
However, $\nu_{\rm QPT}$ for the SU(3) models exhibits a systematic shift toward the trivial value of $\nu_{\rm QPT}=1/D  (D=3)$ as the system size increases.
Therefore, the possibility of a first-order transition remains in the case of SU(3).

\begin{figure}[htb]
  \begin{center}
   \includegraphics[width=0.95\linewidth]{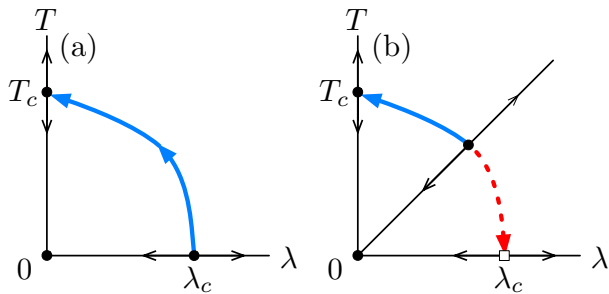}
  \end{center}
\caption{\label{fig7} Schematic phase diagram and renormalization flow. The thick solid (dashed) curves correspond to the second (first) order transition. 
The horizontal axis, $\lambda$, is the coupling ratio of the Heisenberg term, $J$, and the multibody interaction term $Q_m$. 
The open square represents a discontinuous transition. Each solid circle denotes a fixed point, such as the 2D Ising, three-state Potts, and multicritical fixed points. 
The coordination origin corresponds to the low-temperature fixed point. All arrows indicate renormalization flows.
(a) DCP scenario and (b) first-order transition scenario.}
\end{figure}

The nature of the QPT point is important in the discussion of finite-temperature properties,
because it can strongly affect the topology of the thermal phase diagram and also the criticality, as shown in Fig.~\ref{fig7}.  
The SU(N) $JQ_m$ models are expected to exhibit a thermal phase transition if the VBS pattern is characterized by spontaneous symmetry breaking of the lattice.
Thus, consideration of the critical properties of thermal transitions in the vicinity of the QPT point may yield a different perspective on the possibility of DCP occurring in SU(N) $JQ_m$ models.

The universality class of the thermal transition has been discussed for both SU(2) $JQ_2$~\cite{Tsukamoto2009} and $JQ_3$~\cite{JinS2013} models on the square lattice.
The VBS pattern on the square lattice is described by a columnar dimer configuration, which is characterized by the spontaneous breaking of $\pi/2$-rotational symmetry around the center of the plaquette.
Thus, the $Z_4$ symmetry breaking of the VBS order parameter is expected at the critical temperature.
In the 2D case, several models that exhibit $Z_4$ symmetry breaking exist, such as the Ashkin-Teller model~\cite{AKmodel} including the 4-state Potts model~\cite{WuF1982} and the 2D classical XY spin model with the $Z_4$ field (XY+$Z_4$ model). 
In such models, the critical exponent, $\eta$, always satisfies the condition $\eta=1/4$. 
However, the observed exponent $\eta \sim 0.59$ of the SU(2) $JQ_2$ model differs from the expected value~\cite{Tsukamoto2009}. 
In the SU(2) $JQ_2$ model, the VBS order is very weak because the QPT point is located in the vicinity of the limit, 
and the model can only be expressed using the multibody interacting  $Q_m$ term (the dimer limit).
To enhance the VBS order, Jin et al. have focused on the SU(2) $JQ_3$ models~\cite{JinS2013}.
The QMC results they have obtained~\cite{JinS2013} indicate that the criticality is well explained by the Gaussian conformal-field theory with central charge $c=1$;
the thermal exponent, $\nu$, monotonically increases as the system approaches the QPT point, while the following relations between the exponents, $\eta=1/4, \gamma/\nu=7/4$, and $\beta/\nu=1/8$, are retained.
This is a characteristic aspect of the 2D weak Ising universality class~\cite{SuzukiM1974}, and the same behavior has also been observed in the 2D XY+$Z_4$ model~\cite{Jose1977,Rastelli2004a,Rastelli2004b}.
In the case of the classical spin model, $\nu$ monotonically increases as the $Z_4$ symmetry-breaking field, $h_4$, is suppressed and
 finally diverges at the XY limit, where the Kosterliz-Thouless (KT) transition takes place. 
The authors in ref. \cite{JinS2013} have observed that an enhancement of the U(1) symmetry of the VBS order parameter 
is observed at close proximity to the transition temperature and the QPT point,
 when the system size is smaller than a characteristic length scale. 
Since it has been noted that the emergence of additional U(1) symmetry is an important signature of DCP~\cite{SenthilBSVF2005,LouSK2007},
 the numerical result in ref. \cite{JinS2013} is consistent with the presence of a deconfined critical point in the SU(2) $JQ_3$ model.
However,  the observation of U(1) symmetry in the vicinity of the QPT point seems to be natural,
because the $Z_4$ field in the classical model is always marginal at a transition temperature and 
the system becomes the pure XY model at the $h_4 \rightarrow 0$ limit~\cite{Jose1977}.
Thus, the emergence of U(1) symmetry cannot be regarded as sufficient evidence for the presence of a deconfined critical point in this case.
Since the possibility of a first-order transition has been suggested in SU(3) $JQ_2$ model case~\cite{HaradaK2013},  where the same $Z_4$ field is broken,
systematic studies of SU(N) symmetry are necessary.

In contrast to the square-lattice case, the nature of the symmetry-breaking field is different for the honeycomb-lattice case.
When the columnar VBS pattern is characterized by $\pi/3$ rotational symmetry breaking,
the corresponding classical model is expected to be the XY+$Z_3$ model.
Since the $Z_3$ field is relevant in two dimensions, the universality class is explained by the 2D three-state Potts model~\cite{Baxter1982},
 and the emergence of the U(1) symmetry in the VBS order parameter may then be suppressed in the vicinity of the QPT.
Although this is correct in the case of SU(2) spins, the higher SU(N)-symmetric case seems to be controversial. 
The discussion of DCP is based on the noncompact complex projective (NCCP$^{N-1}$) theory with $Z_k$ symmetry-breaking fields~\cite{SenthilVBSF2004,SenthilBSVF2005}.
In this theory, although the $Z_3$ symmetry-breaking fields is relevant, it becomes irrelevant as N increases~\cite{SachdevRAJ1990,SenthilBSVF2004}.
For the SU(2) case, which corresponds to the NCCP$^1$ theory, recent QMC results have indicated that 
the $Z_3$ field is $relevant$ but almost marginal at the QPT~\cite{PujariKDFA2013}.
Therefore, one can expect the first-order transition at the QPT point in the SU(2) case
 and a change of criticality as N increases. 
This indicates that the criticality of the thermal transitions and the topology of the phase diagram are determined base on the order of the QPT. 
If the QPT is continuous, as is expected for larger values of N, and the system approaches the QPT, 
whether or not the universality classes of the thermal transition  are affected is a nontrivial question.

Our previous QMC calculations suggest that the same criticality exists at the QPT regardless of the lattice geometry~\cite{HaradaK2013}.
This implies that the phase diagram topologies are identical in both the square- and the honeycomb-lattice cases.
If one focuses on the most likely and simplest case, two scenarios for the thermal phase diagram can be expected depending on the order of the QPT point: 
(a) The QPT transition is of the second order and the thermal transition is always continuous (Fig.~\ref{fig7}(a)); and
(b) The QPT is a weak first-order transition and the multicritical point exists at a finite temperature (Fig.~\ref{fig7}(b)).
When scenario (b) occurs, we expect to observe crossover behavior and for $\nu$ to change to the trivial value, $\nu=1/D (D=2)$.
From the above discussion, the importance of calculating the thermal phase diagram for different values of N and various lattice geometries with high accuracy is apparent.
Further, such calculations can allow us to consider the possibility of the DCP scenario in the SU(N) $JQ_m$ models.  
Thus, in this paper, we systematically study the thermal phase transitions of the $JQ_2$ model on the square lattice and the $JQ_3$ model on the honeycomb lattice for SU(3) and SU(4) spins.

The layout of this paper is as follows. In Sec. II, we study the thermal transition of the SU(N) $JQ_m$ model.
We begin by introducing the model details and the order parameters evaluated in the QMC computations.
In Sec. III, we present the results of the finite-size scaling analysis for the obtained numerical data.
The criticality of the thermal transition is discussed for the square-lattice and the honeycomb-lattice cases.
Then, we discuss possible scenarios for the QPT of both models from the perspective of the thermal phase diagram.
Finally, we summarize our results in Sec. IV.
%

\section{Models and Method}
\label{sec:Model and Method}
We consider the SU(N) $JQ_{2}$ model on the square lattice and the SU(N) $JQ_{3}$ model on the honeycomb lattice.
Both models are simply expressed by the color-singlet-projection operator, $P_{ij}$, which is defined as 
$P_{ij}=-\frac{1}{N}\sum_{\alpha=1}^{N}\sum_{\beta=1}^{N} S_i^{\alpha \beta}{\bar S}_j^{\beta \alpha}$,
where $S_i^{\alpha \beta}$ is the SU(N) spin generator and ${\bar S}_j^{\beta \alpha}$ is its conjugate.
The model Hamiltonian can be expressed as
\begin{eqnarray}
{\mathcal H}=-J \sum_{( ij )} P_{ij} - Q_2 \sum_{( ij ) ( kl )} P_{ij}P_{kl},
\label{Ham1}
\end{eqnarray}
for the square-lattice case and 
\begin{eqnarray}
{\mathcal H}=-J \sum_{( ij )} P_{ij} - Q_3 \sum_{( ij )( kl )( mn )} P_{ij}P_{kl}P_{mn},
\label{Ham2}
\end{eqnarray}
\begin{figure}[htb]
  \begin{center}
   \includegraphics[width=8cm]{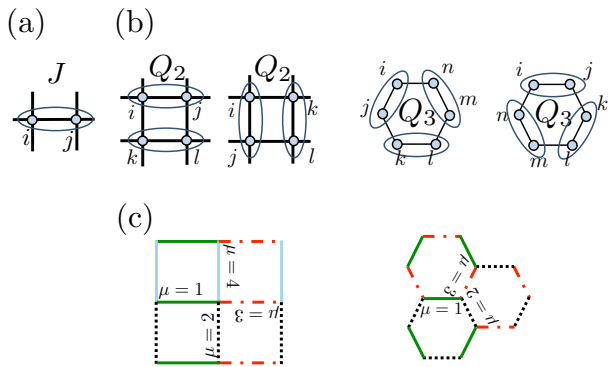}
  \end{center}
\caption{\label{fig1} (Color online) (a) Color-singlet projection operator on a bond. The bold ellipsoids denote a color-singlet dimer state and correspond to $P_{ij}$s. 
(b) Projection operators for $Q_2$ and $Q_3$ terms. (c) Coordination index, $\mu$.}
\end{figure}
for the honeycomb-lattice case, where $(ij)$ indicates the nearest-neighbor sites. 
The summation for the $Q_m$ terms runs over all pairs
without breaking the rotational symmetry of the lattice, as illustrated in Fig. \ref{fig1}.
Since the present lattices are bipartite, the fundamental (conjugate) representation is adapted for the SU(N) spins on A(B) sites.

For the Hamiltonians (\ref{Ham1}) and (\ref{Ham2}), we performed QMC calculations up to $L=256$ for the square- and $L=132$ for the honeycomb-lattice cases, respectively. 
(The number of sites, ${\mathcal N}$, corresponds to ${\mathcal N}=L^2$ and ${\mathcal N}=2L^2$, respectively.)
The QMC code used here is based on the massively parallelized Loop algorithm~\cite{TodoMS2012} provided in the ALPS project code~\cite{ALPS}.
In the computations, we measured the VBS amplitude, which is defined as $\Psi_{\boldsymbol r}  \equiv \sum_{\mu=1}^{z} \exp[ \frac{2\pi i}{z} \mu ] \hat{P}_{\boldsymbol r,r_{\mu}}$,
where $\hat P_{\boldsymbol r,r_\mu}$ is the diagonal component of the projection operator, $z$ is the coordination number of a lattice, and ${\boldsymbol r_\mu}$ represents the neighboring site of ${\boldsymbol r}$ in the $\mu$ direction, respectively (see Fig. \ref{fig1} (b)). 
From $\Psi_{\boldsymbol r}$, the VBS order parameter, which is defined as $\Psi \equiv L^{-2}\sum_{{\boldsymbol r}} \Psi_{\boldsymbol r}$.
After $\Psi_{\boldsymbol r}$ was evaluated, we obtained further quantities: 
the Binder ratio $B_R \equiv \langle \Psi^4 \rangle/ \langle \Psi^2 \rangle ^ 2$; 
the VBS correlation function, $C({\boldsymbol r}) \equiv \langle \Psi_{\boldsymbol r}\Psi_{\boldsymbol 0} \rangle$; 
the correlation ratio $C_R \equiv \frac{C(L/2,L/2)}{C(L/4,L/4)}$; 
the correlation length, $\xi \equiv \frac{1}{|\Delta {\boldsymbol Q}|}\sqrt{ \frac{S({{\boldsymbol Q}_c})}{S(\Delta {\boldsymbol Q})} -1}$; 
and
the static structure factor, $S({\boldsymbol Q})=L^{-2}\sum_{{\boldsymbol r},{\boldsymbol r'}} \exp [-i {\boldsymbol Q}({\boldsymbol r}-{\boldsymbol r}')] \langle \Psi_{\boldsymbol r}\Psi_{\boldsymbol r'} \rangle$.
Here, $\Delta  {\boldsymbol Q}$ denotes the distance between the order wave-vector, ${{\boldsymbol Q}_c}={\boldsymbol 0}$, and the nearest-neighbor positions, $(0,2 \pi /L_y)$ or $(2 \pi /L_x, 0)$.

In this paper, we discuss the thermal transition criticality by changing the coupling constants, $J$ and $Q_m$.
It is convenient to introduce a length scale associated with the distance from the QPT point,
where the ground state changes from the N\'eel state to the VBS state.
The QPT points were previously evaluated in ref. ~\cite{HaradaK2013} and are summarized in table \ref{table1}.
The coupling ratio, $\lambda=J/(J+Q_m)$, of the QPT point depends strongly on the lattice geometry, and also on the degree of freedom of the SU(N) spin.
Therefore, we introduce a normalized coupling constant that is defined as $\Lambda=\lambda/\lambda_{c}$, where $\lambda_{c}$ is the critical value at the QPT point.
From this definition, one can easily see that $\Lambda=0$ and $1$ correspond to the dimer limit and the QPT point, respectively.
\begin{table}
\begin{center}
  \begin{tabular}{ccc}
  \hline \hline 
  SU(N) & $JQ_2$ & $JQ_3$\\ \hline 
  2 & $\lambda_c=0.042$ & $\lambda_c=0.456$\\
  3 & $\lambda_c=0.665$ & $\lambda_c=0.796$\\
  4 & $\lambda_c=0.917$ & $\lambda_c=0.985$\\ \hline \hline
  \end{tabular}
  \caption{\label{table1} Critical points of SU(N) $JQ_m$ models. $\lambda_c$ is the critical value of the coupling ratio defined as $\lambda \equiv J/(J+Q_m)$, where  $J$ and $Q_m$ are the coupling constants. All values are given in ref.~\cite{HaradaK2013}. }
\end{center}
\end{table}
%

\section{Numerical Results and Finite-Size Scaling Analysis}
\label{ResultsDiscussion}
%

\begin{figure}[htb]
  \begin{center}
    \includegraphics[width=0.6\linewidth]{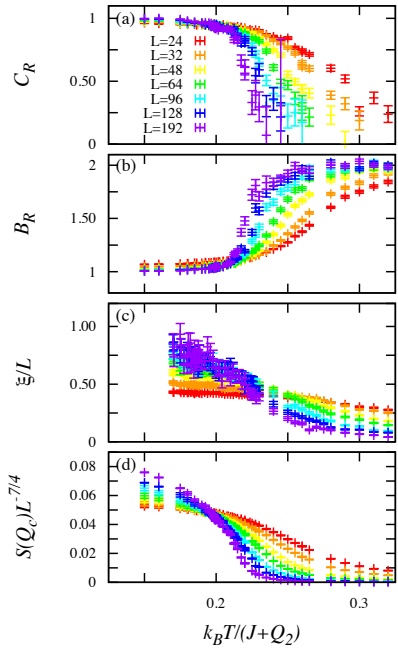}
  \end{center}
  \vspace{-0.5cm}
  \caption{\label{fig2a} (Color online) Temperature dependence of $C_R$, $B_R$, $\xi/L$, and $S({\boldsymbol Q}_c)L^{-\frac{\gamma}{\nu}}$ in the SU(3) square-lattice model at $\Lambda=0.5$. }
\end{figure}

\begin{figure}
  \begin{center}
    \includegraphics[width=0.6\linewidth]{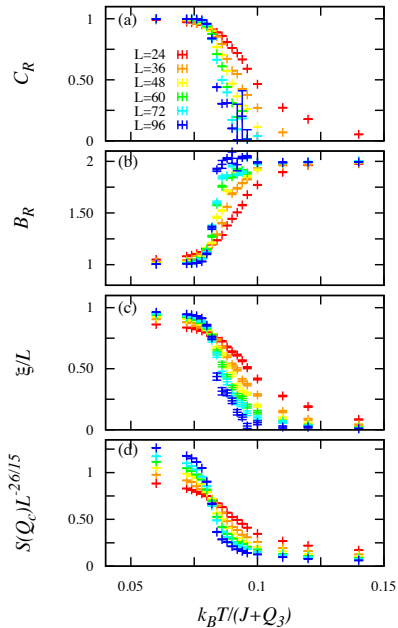}
  \end{center}
  \vspace{-0.5cm}
  \caption{\label{fig2b} (Color online) Temperature dependence of $C_R$, $B_R$, $\xi/L$, and $S({\boldsymbol Q}_c)L^{-\frac{\gamma}{\nu}}$  for the SU(4) honeycomb-lattice model at $\Lambda=0.5$.}
\end{figure}

In Figs. \ref{fig2a} and \ref{fig2b}, we show the temperature dependence of $C_R$, $B_R$, $\xi$, and $S({\boldsymbol Q}_c)$ at $\Lambda=0.5$, 
which is the middle distance between the QPT point and the dimer limit.
Since clear crosses are always observed for $0 \le \Lambda \lesssim 1$ as the temperature decreases, 
the thermal transition from the paramagnetic to the VBS phase is expected to be of the second order.

\begin{figure}[htb]
  \begin{center}
   \includegraphics[width=0.7\linewidth]{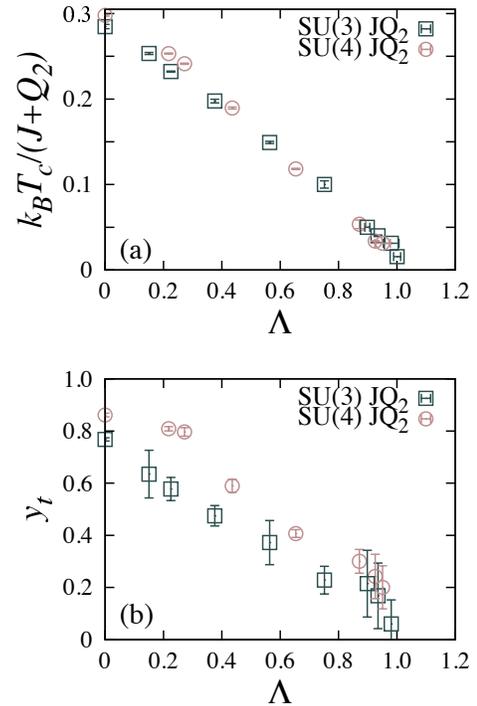}
  \end{center}
   \vspace{-0.7cm}
\caption{\label{fig3} (Color online) (a) Critical temperature and (b) renormalization group eigenvalue, $y_t$, for temperature in the square-lattice case. 
The open squares (circles) are the SU(3) (SU(4)) results. $y_t$ is estimated by extrapolation to the thermodynamic limit,
$\Lambda=0$ corresponds to the dimer limit, where $J=0$, and $\Lambda=1$ is the QPT point.}
\end{figure}
\begin{figure}[htb]
  \begin{center}
   \includegraphics[width=\linewidth]{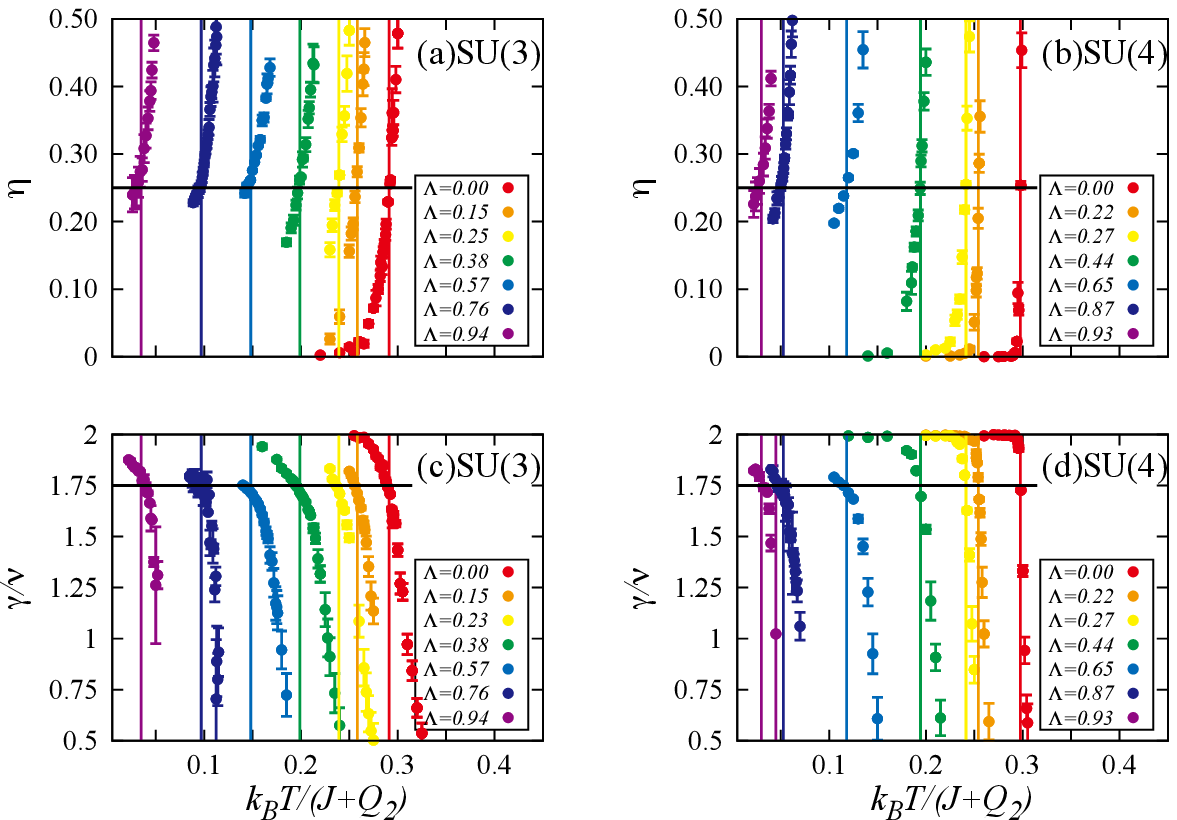}
  \end{center}
    \vspace{-0.7cm}
\caption{\label{fig4} (Color online) $\Lambda$ dependence of effective $\eta$ and $\gamma/\nu$ of SU(N) $JQ_2$ models.
All values were evaluated from the assumptions, $C(R=L/\sqrt{2})|_{T\sim T_c} \sim L^{-\eta}$ and $S({\boldsymbol Q}_c)|_{T\sim T_c} \sim L^{\frac{\gamma}{\nu}}$ , 
which are approximately satisfied in the vicinity of the critical temperatures. 
The vertical colored lines are critical temperatures and the black horizontal lines correspond to 
the values of the exponents for the 2D Ising universality class ($\eta=1/4$ and $\gamma/\nu=7/4$).}
\end{figure}

\begin{figure}[htb]
  \begin{center}
   \includegraphics[width=0.95\linewidth]{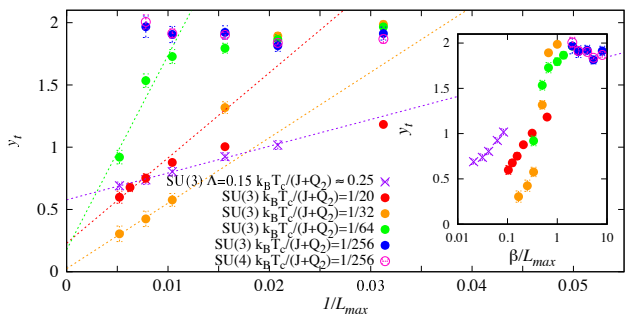}
  \end{center}
    \vspace{-0.7cm}
\caption{\label{fig11} (Color online) System-size dependence of $y_t$ estimated from $B_R$. All $y_t$ values were evaluated from the Bayesian scaling analysis for several data sets labeled $L_{{\rm max}}$ (see text). 
The extrapolated values are expected to be those in the thermodynamic limit. The dotted line is a guide for the eye. The inset is the same result plotted as a function of $\beta/L_{{\rm max}}$.}
\end{figure}

To discuss the universality class, we performed a FSS analysis of 
$\xi$, $C_R$, $B_R$, and $S({\boldsymbol Q}_c)$, assuming the scaling forms, $\xi/L \sim g_{\xi}[ L^{y_t}(T-T_c) ]$, $C_R \sim g_{C_R}[L^{y_t}(T-T_c)]$, 
$B_R \sim g_{B_R}[L^{y_t}(T-T_c)]$, and $S({\boldsymbol Q}_c)L^{-\frac{\gamma}{\nu}} \sim g_{S_Q}[L^{y_t}(T-T_c)]$, where $y_t={\nu}^{-1}$ and $g_X[x]$ is a scaling function.
We applied the Bayesian scaling analysis~\cite{HaradaKBayes2011} to the FSS analysis of larger system size sets and estimated the values as follows.
First, the critical temperature, $T_c$, and $y_t$ were evaluated from $\xi$ and $C_R$ (or $\xi$ and the Binder ratio $B_R$), 
because their scaling forms contain only two variables, $T_c$ and $y_t$. 
Both $T_c$ and $y_t$ were optimized simultaneously from the $\xi$ and $C_R$ data set.
In detail, we evaluated $T_c$ and $y_t$ for several data sets  labeled $L_{{\rm max}}$ that include four different system sizes, for example, $L_{{\rm max}}=128$ includes $L=\{48,64,96,128\}$, $L_{{\rm max}}=96$ includes $L=\{48,60,72,96\}$, and so on.
Since apparent system-size dependence is observed for $\Lambda > 0$, we evaluated the extrapolated values of $T_c$ and $y_t$ in the limit $L_{{\rm max}} \rightarrow \infty$ from the large-system sets.
(One example of this size dependence is the result at $\Lambda=0.15$ for SU(3) shown in Fig. \ref{fig11}.)
After we obtained $T_c$ and $y_t$ for the thermodynamic limit, $\eta$ and $\gamma/\nu$ were independently from the correlation function, 
$C({\boldsymbol r})$, and $S({\boldsymbol Q}_c)$. 

We summarize the estimated $y_t(=\nu^{-1})$ and $T_c$ in Fig. \ref{fig3} for the square-lattice case (the SU(N) $JQ_2$ model). 
In the square-lattice model, $y_t$ ($\nu$) monotonically decreases (increases) as the system approaches the quantum critical point, in both the SU(3) and the SU(4) cases.
In contrast to $y_t$, we observe that $\eta$ and $\gamma/\nu$ take constant values for $\Lambda<0.97$.
Figures \ref{fig4} (a) and (b) show the $\Lambda$ dependence of the effective $\eta$ estimated from the assumption that $C(R) \sim R^{-\eta}$. 
From Figs. \ref{fig4} (a) and (b), it is apparent that $\eta$ clearly crosses $\eta=1/4$ at critical temperatures within the error bars.
 In the same manner, the effective $\gamma/\nu$ is estimated from the form, $S({\boldsymbol Q}_c=0) \sim L^{\gamma/\nu}$.
Figures \ref{fig4} (c) and (d) present $\gamma/\nu$ evaluated from the data for $L \ge 96$.
We can confirm from Fig. \ref{fig4} that $\gamma/\nu$ crosses the value $7/4$ at critical temperatures. 
Thus we conclude that $\eta$ and $\gamma/\nu$ satisfy $\eta=1/4$ and $\gamma/\nu=7/4$ at critical temperatures, within the error bars.
The obtained exponents are the same as those of the 2D Ising universality class. 
$y_t$ ($\nu$) itself varies depending on $\Lambda$, but the other exponents, such as $\eta$ and $\gamma/\nu$, are constant. 
This behavior is known as the 2D Ising weak universality~\cite{SuzukiM1974} and is consistent with the results reported in ref. \cite{JinS2013} for the SU(2) $JQ_3$ model.

To approach the QPT point from the finite temperature region, we performed these calculations 
at very low fixed temperatures by varying $\lambda$.
With limited system size, we observed an apparent increase in $y_t$. 
However, as we discuss below, this is due to crossover from the mean-field type behavior to 
the true asymptotic behavior, and should not be taken as an evidence suggesting a first-order transition.
(This is a slightly confusing point since the mean-field value for $y_t=2$ happens to be equal to 
the expected value for the first-order transition in two dimensions.)
Figure \ref{fig11} shows the system size dependence of $y_t$ at $k_BT_c/(J+Q_2)=1/20$, $1/32$, $1/64$, and $1/256$ for the SU(3) case.
Each value of $y_t$ is the FSS result of $B_R$ for several data sets labeled $L_{{\rm max}}$ that include three different system sizes; for example, in the square-lattice case, $L_{{\rm max}}=48$ includes $L=\{24,32,48\}$, $L_{{\rm max}}=72$ contains $L=\{48,64,72\}$ and so on.
At $k_BT_c/(J+Q_2)=1/20$, $y_t$ $systematically$ decreases as $L_{{\rm max}}$ increases and takes the approximate value $y_t \sim 0.23$ in the thermodynamic limit.
However, in the lower temperature region, we observe that $y_t$ exhibits a crossover from the mean-field value;
 the data for small $L_{{\rm max}}$ indicate $y_t \sim 2 (=\frac{1}{\nu})$, but $y_t$ decreases suddenly when the system size becomes larger than a characteristic length, $L_c$.
In the case of $k_BT_c/(J+Q_2)=1/32$ and $1/64$,  we estimated $L_c \sim 72$ and $L_c \sim 192$, respectively.
However, when $k_BT_c/(J+Q_2)=1/256$,  we obtained a data with the exponents of the mean-field value in both the SU(3) and SU(4) case.
Therefore, we can obtain the correct values from the data for $L>L_c$ in the FSS analysis, 
 while we estimate the mean-field values from the $L<L_c$ data.
 This $L_c$ is natively related to the development of the correlation length along the imaginary-time direction, $\xi_{\tau}$; 
 the thermal criticality can be observed after $\xi_{\tau}$ approximately exceeds the inverse temperature, $\beta$. 
 (It is expected that $L_c \sim \xi_{\tau} \sim a \beta$, where $a$ is an unknown constant.)  
 In the present case, the correlation along the real space direction is well developed for $\xi_{\tau} < \beta$. 
 Thus, the system can be described by an effective model with long-range interactions.
Similar crossover is observed for the critical exponent $y_t$ $(=1/\nu)$ in the 2D Ising models with long-range interactions~\cite{Luijten1997}.
In the Ising model, $y_t$ depends on the ratio between the interaction range and system size. 
 When the interacting range is significantly larger than the system size, mean-field-type behavior is observed.
From the extrapolated results for $y_t$, it can be stated that the universality class of the thermal transition for $T_c \ge 1/64$
is explained by that of the 2D classical XY+$Z_4$ model, and is therefore the weak 2D Ising universality class.

\begin{figure}[htb]
  \begin{center}
   \includegraphics[width=0.7\linewidth]{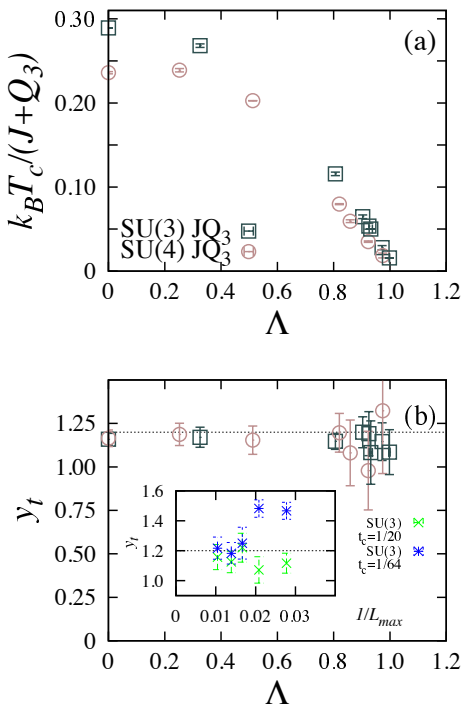}
  \end{center}
   \vspace{-0.7cm}
\caption{\label{fig5} (Color online) (a) Critical temperature and (b) renormalization group eigenvalue for temperature in the honeycomb-lattice case.
$y_t$ is also evaluated from extrapolation to the thermodynamic limit. 
The inset of (b) is the system-size dependence of $y_t$ for the results obtained from the fixed-temperature calculations  at $k_BT_c/(J+Q_3)=1/20$ and $1/64$ (see in text).
The label $t_c$ in the figure means $k_BT_c/(J+Q_3)$.
The open squares (circles) are the results for the SU(3) (SU(4)) spins.  The black dotted line is the value of the 2D three-state Potts case, $y_t=6/5$.  }
\end{figure}
\begin{figure}[htb]
  \begin{center}
   \includegraphics[width=\linewidth]{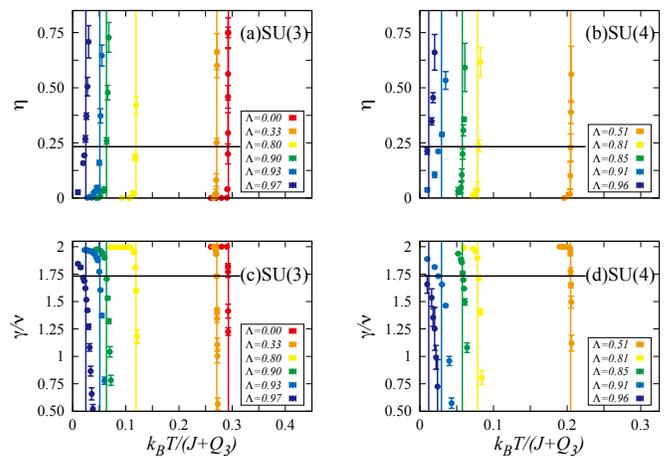}
  \end{center}
   \vspace{-0.7cm}
\caption{\label{fig6} (Color online) $\Lambda$ dependence of estimated $\eta$ and $\gamma/\nu$ for $JQ_3$ model on honeycomb lattice. 
(a) and (b) ((c) and (d)) are $\eta$ ($\gamma/\nu$) results for the SU(3) and SU(4) cases, respectively.
The vertical colored lines denote critical temperatures and the black horizontal lines are critical exponents for the 2D three-state Potts model, $\eta=4/15$ and $\gamma/\nu=26/15$.}
\end{figure}
\begin{figure}[htb]
  \begin{center}
   \includegraphics[width=0.95\linewidth]{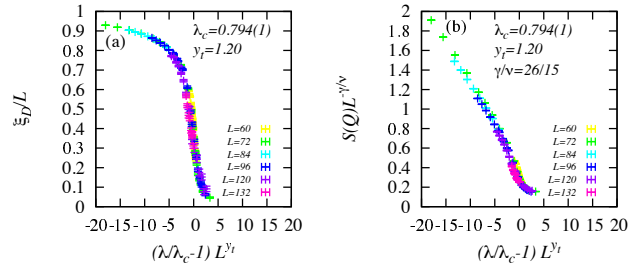}
  \end{center}
   \vspace{-0.7cm}
\caption{\label{fig10} (Color online) Finite-size scaling analysis for SU(3) honeycomb-lattice case at $k_BT_c/(J+Q_3)=1/64$. (a) Correlation length and (b) static structure factor. }
\end{figure}

Next, we focus on the criticality of the SU(N) $JQ_3$ model on the honeycomb lattice. 
In Fig. \ref{fig5}, we summarize $y_t$=$\nu^{-1}$ and $T_c$ for the SU(3) and SU(4) $JQ_3$ model.
We find that $y_t$=6/5 ($\nu=5/6$) is well satisfied even in the vicinity of the QPT limit of $\Lambda=1$
and that the size dependence of $y_t$ is quite small for $\Lambda \lesssim 0.95$. 
This value is consistent with that of the 2D three-state Potts universality.
In Fig. \ref{fig6}, we show $\eta$ and $\gamma/\nu$ that were estimated in the same manner as in the square-lattice case. 
The 2D three-state Potts universality is also confirmed directly; $\eta=4/15$ and $\gamma/\nu=26/15$ are satisfied at the critical temperatures within error bars.
The present columnar VBS pattern is characterized by the $\pi/3$-rotational symmetry breaking, reflecting the honeycomb-lattice background. 
Thus, it is expected that the related classical model with the same universality class is the 2D XY+$Z_3$ model.
Since the $Z_3$ field is strongly relevant in two dimensions, the exponents are not affected by the coupling constants $J$ and $Q_3$.

The fact that the $Z_3$ field is relevant may help us discuss the possibility of DCP occurring.
If the present honeycomb $JQ_3$ model can be well mapped onto the 2D three-state Potts model 
and the change in the coupling ratio can be regarded as the variation of certain parameters, for example, the transverse field in the conventional 2D Ising model,
the criticality in the QPT limit is explained by the 3D three-state Potts model.
In that case, the QPT should exhibit a weak first-order transition~\cite{WuFY3D3Potts,WuFY3D3Potts2,BloteHWJ1979,JankeW1997} 
and the first-order transition line should extend in the finite-temperature region. The length of the first-order transition line may be finite but is too short to be observed (see Fig.~\ref{fig7} (b)).
This means that the value of $\nu$ should change from the 2D three-state Potts value to the trivial value of $\nu=1/D$ $(D=2)$ via the value at the multicritical fixed point. 
However, such crossover behavior is not observed when the system approaches the QPT point. 
We also performed the fixed-temperature calculations for the honeycomb-lattice case. 
When we vary $\lambda$ for fixed $k_BT/(J+Q_3)=1/64$, where the critical point corresponds to $\Lambda \sim 0.99$, the 2D three-state Potts universality is still observed.
Figure \ref{fig10} shows the FSS results for the SU(3) case. 
We obtain data collapse for $L>80$ if we set the critical exponents to those of the 2D three-state Potts universality.
In the case of the honeycomb-lattice model, we expect that the crossover behavior from the mean-field theory exists for $L<80$, 
but it is very weak. Therefore, it is difficult to identify the conventional system-size dependence. 
This result indicates that the development of $L_c$ is relatively suppressed in comparison with the square-lattice case at the same temperature.

The obtained thermal phase diagram for $\Lambda \lesssim 0.99$ supports the possibility of scenario (a) in Fig. \ref{fig7}, 
because it seems unlikely that $\nu$ will approach the trivial value of $1/D$ in both the square-lattice and the honeycomb-lattice cases.
If the scenario (b) occurs, the multicritical point should exist at quite a low temperature, i.e., $k_BT/(J+Q_m) < O(10^{-2})$.
This is still consistent with our previous discussion of the QPT point~\cite{HaradaK2013};
a systematic increase in $\nu_{\rm QPT}$ towards the trivial value is observed for $L>128$ in the SU(3) square-lattice model.
If the dynamical exponent for the DCP is unity, $k_BT/(J+Q_m) < O(10^{-2})$ corresponds to the length scale, $L > O(10^2)$.
This implies that the correlation length is very large and almost diverging.

\section{Summary}
\label{Summary}
In this paper, we have investigated the thermal transitions of $JQ_2$ models on the square lattice and $JQ_3$ models 
on the honeycomb lattice for SU(3) and SU(4) spins. 
We have found that the criticality of the SU(N) square-lattice model is well explained by the 2D weak Ising universality class in both the SU(3) and SU(4) cases, which is in agreement with Jin and Sandvik's result~\cite{JinS2013} for the SU(2) $JQ_3$ model.
The thermal exponent, $\nu$, monotonically increases as the system approaches the QPT limit, 
and the decrease in $\nu$ that should occur if $\nu$ eventually reaches its first-order transition value of $1/D$ has not been observed.
Thus, the first-order transition appears to be less likely for $k_BT_c/(J+Q_m) > O(10^{-2})$.
In the honeycomb-lattice case, reflecting the fact that the $Z_3$ field is strongly relevant, $\nu$ always exhibits the 2D three-state Potts value. 
From the obtained results, we have discussed possible scenarios for the thermal phase diagram.  
If the first-order transition occurs, we may observe critical behaviors with strong system-size corrections.
However for $k_BT_c/(J+Q_m)>1/64$, cross-over behavior is not observed clearly in our results.  
To determine the thermal phase diagram (a) or (b) occurring in the present models, 
the numerical calculations for extremely large system sizes are required, because the drastic development of $L_c$ is expected in the vicinity of the QPT.

\section*{Acknowledgments}
\label{ackno}
We thank T. Okubo for fruitful discussions.  This work is supported by MEXT Grand-in-Aid for Scientific Research (b) (25287097) and Scientific Research (c) (26400392). 
We are grateful for use of the computational resources of the K computer in the RIKEN Advanced Institute for Computational Science 
through the HPCI System Research project (Project ID: hp120283 and hp130081).
We also thank numerical resources in the ISSP Supercomputer Center at University of Tokyo, 
and the Research Center for Nano-micro Structure Science and Engineering at University of Hyogo.


\begin{thebibliography}{32}
\expandafter\ifx\csname natexlab\endcsname\relax\def\natexlab#1{#1}\fi
\expandafter\ifx\csname bibnamefont\endcsname\relax
  \def\bibnamefont#1{#1}\fi
\expandafter\ifx\csname bibfnamefont\endcsname\relax
  \def\bibfnamefont#1{#1}\fi
\expandafter\ifx\csname citenamefont\endcsname\relax
  \def\citenamefont#1{#1}\fi
\expandafter\ifx\csname url\endcsname\relax
  \def\url#1{\texttt{#1}}\fi
\expandafter\ifx\csname urlprefix\endcsname\relax\def\urlprefix{URL }\fi
\providecommand{\bibinfo}[2]{#2}
\providecommand{\eprint}[2][]{\url{#2}}

\bibitem[{\citenamefont{Ginzburg and Landau}(1950)}]{GL}
\bibinfo{author}{\bibfnamefont{V.}~\bibnamefont{Ginzburg}} \bibnamefont{and}
  \bibinfo{author}{\bibfnamefont{L.}~\bibnamefont{Landau}},
  \bibinfo{journal}{Zh. Eksp. Teor. Fiz.} \textbf{\bibinfo{volume}{20}},
  \bibinfo{pages}{1064} (\bibinfo{year}{1950}).

\bibitem[{\citenamefont{Wilson and Kogut}(1974)}]{GLW}
\bibinfo{author}{\bibfnamefont{K.~G.} \bibnamefont{Wilson}} \bibnamefont{and}
  \bibinfo{author}{\bibfnamefont{J.}~\bibnamefont{Kogut}},
  \bibinfo{journal}{Phys. Rep.} \textbf{\bibinfo{volume}{12}},
  \bibinfo{pages}{75} (\bibinfo{year}{1974}).

\bibitem[{\citenamefont{Senthil
  et~al.}(2004{\natexlab{a}})\citenamefont{Senthil, Vishwanath, Balentz,
  Sachdev, and Fisher}}]{SenthilVBSF2004}
\bibinfo{author}{\bibfnamefont{T.}~\bibnamefont{Senthil}},
  \bibinfo{author}{\bibfnamefont{A.}~\bibnamefont{Vishwanath}},
  \bibinfo{author}{\bibfnamefont{L.}~\bibnamefont{Balentz}},
  \bibinfo{author}{\bibfnamefont{S.}~\bibnamefont{Sachdev}}, \bibnamefont{and}
  \bibinfo{author}{\bibfnamefont{M.~P.~A.} \bibnamefont{Fisher}},
  \bibinfo{journal}{Science} \textbf{\bibinfo{volume}{303}},
  \bibinfo{pages}{1490} (\bibinfo{year}{2004}{\natexlab{a}}).

\bibitem[{\citenamefont{Senthil
  et~al.}(2004{\natexlab{b}})\citenamefont{Senthil, Balents, Sachdev,
  Vishwanath, and Fisher}}]{SenthilBSVF2004}
\bibinfo{author}{\bibfnamefont{T.}~\bibnamefont{Senthil}},
  \bibinfo{author}{\bibfnamefont{L.}~\bibnamefont{Balents}},
  \bibinfo{author}{\bibfnamefont{S.}~\bibnamefont{Sachdev}},
  \bibinfo{author}{\bibfnamefont{A.}~\bibnamefont{Vishwanath}},
  \bibnamefont{and} \bibinfo{author}{\bibfnamefont{M.~P.~A.}
  \bibnamefont{Fisher}}, \bibinfo{journal}{Phys. Rev. B}
  \textbf{\bibinfo{volume}{70}}, \bibinfo{pages}{144407}
  (\bibinfo{year}{2004}{\natexlab{b}}).

\bibitem[{\citenamefont{Senthil et~al.}(2005)\citenamefont{Senthil, Balents,
  Sachdev, Vishwanath, and Fisher}}]{SenthilBSVF2005}
\bibinfo{author}{\bibfnamefont{T.}~\bibnamefont{Senthil}},
  \bibinfo{author}{\bibfnamefont{L.}~\bibnamefont{Balents}},
  \bibinfo{author}{\bibfnamefont{S.}~\bibnamefont{Sachdev}},
  \bibinfo{author}{\bibfnamefont{A.}~\bibnamefont{Vishwanath}},
  \bibnamefont{and} \bibinfo{author}{\bibfnamefont{M.~P.~A.}
  \bibnamefont{Fisher}}, \bibinfo{journal}{J. Phys. Soc. Jpn.}
  \textbf{\bibinfo{volume}{74 Suppl.}}, \bibinfo{pages}{1}
  (\bibinfo{year}{2005}).

\bibitem[{\citenamefont{Sandvik}(2007)}]{Sandvik2007}
\bibinfo{author}{\bibfnamefont{A.~W.} \bibnamefont{Sandvik}},
  \bibinfo{journal}{Phys.\ Rev.\ Lett.} \textbf{\bibinfo{volume}{98}},
  \bibinfo{pages}{227202} (\bibinfo{year}{2007}).

\bibitem[{\citenamefont{Melko and Kaul}(2008)}]{MelkoRG2008}
\bibinfo{author}{\bibfnamefont{R.~G.} \bibnamefont{Melko}} \bibnamefont{and}
  \bibinfo{author}{\bibfnamefont{R.~K.} \bibnamefont{Kaul}},
  \bibinfo{journal}{Phys. Rev. Lett.} \textbf{\bibinfo{volume}{100}},
  \bibinfo{pages}{017203} (\bibinfo{year}{2008}).

\bibitem[{\citenamefont{Kuklov et~al.}(2008)\citenamefont{Kuklov, Matsumoto,
  Prokof'ev, Svistunov, and Troyer}}]{KuklovAB2008}
\bibinfo{author}{\bibfnamefont{A.~B.} \bibnamefont{Kuklov}},
  \bibinfo{author}{\bibfnamefont{M.}~\bibnamefont{Matsumoto}},
  \bibinfo{author}{\bibfnamefont{N.~V.} \bibnamefont{Prokof'ev}},
  \bibinfo{author}{\bibfnamefont{B.~V.} \bibnamefont{Svistunov}},
  \bibnamefont{and} \bibinfo{author}{\bibfnamefont{M.}~\bibnamefont{Troyer}},
  \bibinfo{journal}{Phys. Rev. Lett.} \textbf{\bibinfo{volume}{101}},
  \bibinfo{pages}{050405} (\bibinfo{year}{2008}).

\bibitem[{\citenamefont{Lou et~al.}(2009)\citenamefont{Lou, Sandvik, and
  Kawashima}}]{LouSK2007}
\bibinfo{author}{\bibfnamefont{J.}~\bibnamefont{Lou}},
  \bibinfo{author}{\bibfnamefont{A.~W.} \bibnamefont{Sandvik}},
  \bibnamefont{and}
  \bibinfo{author}{\bibfnamefont{N.}~\bibnamefont{Kawashima}},
  \bibinfo{journal}{Phys.\ Rev.\ B} \textbf{\bibinfo{volume}{80}},
  \bibinfo{pages}{180414} (\bibinfo{year}{2009}).
  
\bibitem[{\citenamefont{Sandvik}(2010)}]{Sandvik2010}
\bibinfo{author}{\bibfnamefont{A.~W.} \bibnamefont{Sandvik}},
  \bibinfo{journal}{Phys. Rev. Lett.} \textbf{\bibinfo{volume}{104}},
  \bibinfo{pages}{177201} (\bibinfo{year}{2010}).

\bibitem[{\citenamefont{Kaul and Sandvik}(2012)}]{KaulS2012}
\bibinfo{author}{\bibfnamefont{R.~K.} \bibnamefont{Kaul}} \bibnamefont{and}
  \bibinfo{author}{\bibfnamefont{A.~W.} \bibnamefont{Sandvik}},
  \bibinfo{journal}{Phys.\ Rev.\ Lett.} \textbf{\bibinfo{volume}{108}},
  \bibinfo{pages}{137201} (\bibinfo{year}{2012}).

\bibitem[{\citenamefont{Chen et~al.}(2013)\citenamefont{Chen, Huang, Deng,
  Kuklov, Prokof'ev, and Svistunov}}]{ChenK2013}
\bibinfo{author}{\bibfnamefont{K.}~\bibnamefont{Chen}},
  \bibinfo{author}{\bibfnamefont{Y.}~\bibnamefont{Huang}},
  \bibinfo{author}{\bibfnamefont{Y.}~\bibnamefont{Deng}},
  \bibinfo{author}{\bibfnamefont{A.~B.} \bibnamefont{Kuklov}},
  \bibinfo{author}{\bibfnamefont{N.~V.} \bibnamefont{Prokof'ev}},
  \bibnamefont{and} \bibinfo{author}{\bibfnamefont{B.~V.}
  \bibnamefont{Svistunov}}, \bibinfo{journal}{Phys. Rev. Lett.}
  \textbf{\bibinfo{volume}{110}}, \bibinfo{pages}{185701}
  (\bibinfo{year}{2013}).

\bibitem[{\citenamefont{Harada et~al.}(2013)\citenamefont{Harada, Suzuki,
  Okubo, Matsuo, Lou, Watanabe, Todo, and Kawashima}}]{HaradaK2013}
\bibinfo{author}{\bibfnamefont{K.}~\bibnamefont{Harada}},
  \bibinfo{author}{\bibfnamefont{T.}~\bibnamefont{Suzuki}},
  \bibinfo{author}{\bibfnamefont{T.}~\bibnamefont{Okubo}},
  \bibinfo{author}{\bibfnamefont{H.}~\bibnamefont{Matsuo}},
  \bibinfo{author}{\bibfnamefont{J.}~\bibnamefont{Lou}},
  \bibinfo{author}{\bibfnamefont{H.}~\bibnamefont{Watanabe}},
  \bibinfo{author}{\bibfnamefont{S.}~\bibnamefont{Todo}}, \bibnamefont{and}
  \bibinfo{author}{\bibfnamefont{N.}~\bibnamefont{Kawashima}},
  \bibinfo{journal}{Phys. Rev. B} \textbf{\bibinfo{volume}{88}},
  \bibinfo{pages}{220408(R)} (\bibinfo{year}{2013}).

\bibitem[{\citenamefont{Pujari et~al.}(2013)\citenamefont{Pujari, Damle, and
  Alet}}]{PujariKDFA2013}
\bibinfo{author}{\bibfnamefont{S.}~\bibnamefont{Pujari}},
  \bibinfo{author}{\bibfnamefont{K.}~\bibnamefont{Damle}}, \bibnamefont{and}
  \bibinfo{author}{\bibfnamefont{F.}~\bibnamefont{Alet}},
  \bibinfo{journal}{Phys. Rev. Lett.} \textbf{\bibinfo{volume}{111}},
  \bibinfo{pages}{087203} (\bibinfo{year}{2013}).

\bibitem[{\citenamefont{Tsukamoto et~al.}(2009)\citenamefont{Tsukamoto, Harada,
  and Kawashima}}]{Tsukamoto2009}
\bibinfo{author}{\bibfnamefont{M.}~\bibnamefont{Tsukamoto}},
  \bibinfo{author}{\bibfnamefont{K.}~\bibnamefont{Harada}}, \bibnamefont{and}
  \bibinfo{author}{\bibfnamefont{N.}~\bibnamefont{Kawashima}},
  \bibinfo{journal}{Journal of Physics: Conf. Ser.}
  \textbf{\bibinfo{volume}{150}}, \bibinfo{pages}{042218}
  (\bibinfo{year}{2009}).

\bibitem[{\citenamefont{Jin and Sandvik}(2013)}]{JinS2013}
\bibinfo{author}{\bibfnamefont{S.}~\bibnamefont{Jin}} \bibnamefont{and}
  \bibinfo{author}{\bibfnamefont{A.~W.} \bibnamefont{Sandvik}},
  \bibinfo{journal}{Phys. Rev. B} \textbf{\bibinfo{volume}{87}},
  \bibinfo{pages}{180404(R)} (\bibinfo{year}{2013}).

\bibitem[{\citenamefont{Ashkin and Teller}(1943)}]{AKmodel}
\bibinfo{author}{\bibfnamefont{J.}~\bibnamefont{Ashkin}} \bibnamefont{and}
  \bibinfo{author}{\bibfnamefont{E.}~\bibnamefont{Teller}},
  \bibinfo{journal}{Phys. Rev.} \textbf{\bibinfo{volume}{64}},
  \bibinfo{pages}{2542} (\bibinfo{year}{1943}).

\bibitem[{\citenamefont{Wu}(1982{\natexlab{a}})}]{WuF1982}
\bibinfo{author}{\bibfnamefont{F.}~\bibnamefont{Wu}}, \bibinfo{journal}{Rev.
  Mod. Phys.} \textbf{\bibinfo{volume}{54}}, \bibinfo{pages}{235}
  (\bibinfo{year}{1982}{\natexlab{a}}).

\bibitem[{\citenamefont{Suzuki}(1974)}]{SuzukiM1974}
\bibinfo{author}{\bibfnamefont{M.}~\bibnamefont{Suzuki}},
  \bibinfo{journal}{Prog. Theor. Phys.} \textbf{\bibinfo{volume}{51}},
  \bibinfo{pages}{1992} (\bibinfo{year}{1974}).

\bibitem[{\citenamefont{Jos\'e et~al.}(1977)\citenamefont{Jos\'e, Kadanoff,
  Kirkpatrick, and Nelson}}]{Jose1977}
\bibinfo{author}{\bibfnamefont{J.~V.} \bibnamefont{Jos\'e}},
  \bibinfo{author}{\bibfnamefont{L.~P.} \bibnamefont{Kadanoff}},
  \bibinfo{author}{\bibfnamefont{S.}~\bibnamefont{Kirkpatrick}},
  \bibnamefont{and} \bibinfo{author}{\bibfnamefont{D.~R.}
  \bibnamefont{Nelson}}, \bibinfo{journal}{Phys. Rev. B}
  \textbf{\bibinfo{volume}{16}}, \bibinfo{pages}{1217} (\bibinfo{year}{1977}).

\bibitem[{\citenamefont{Rastelli
  et~al.}(2004{\natexlab{a}})\citenamefont{Rastelli, Regina, and
  Tassi}}]{Rastelli2004a}
\bibinfo{author}{\bibfnamefont{E.}~\bibnamefont{Rastelli}},
  \bibinfo{author}{\bibfnamefont{S.}~\bibnamefont{Regina}}, \bibnamefont{and}
  \bibinfo{author}{\bibfnamefont{A.}~\bibnamefont{Tassi}},
  \bibinfo{journal}{Phys. Rev. B} \textbf{\bibinfo{volume}{69}},
  \bibinfo{pages}{174407} (\bibinfo{year}{2004}{\natexlab{a}}).

\bibitem[{\citenamefont{Rastelli
  et~al.}(2004{\natexlab{b}})\citenamefont{Rastelli, Regina, and
  Tassi}}]{Rastelli2004b}
\bibinfo{author}{\bibfnamefont{E.}~\bibnamefont{Rastelli}},
  \bibinfo{author}{\bibfnamefont{S.}~\bibnamefont{Regina}}, \bibnamefont{and}
  \bibinfo{author}{\bibfnamefont{A.}~\bibnamefont{Tassi}},
  \bibinfo{journal}{Phys. Rev. B} \textbf{\bibinfo{volume}{70}},
  \bibinfo{pages}{174447} (\bibinfo{year}{2004}{\natexlab{b}}).

\bibitem[{\citenamefont{Baxter}(1982)}]{Baxter1982}
\bibinfo{author}{\bibfnamefont{R.~J.} \bibnamefont{Baxter}},
  \emph{\bibinfo{title}{Exactly solved models in statistical mechanics}}
  (\bibinfo{publisher}{Academic Press Inc.}, \bibinfo{address}{London},
  \bibinfo{year}{1982}).

\bibitem[{\citenamefont{Sachdev and Jalabert}(1990)}]{SachdevRAJ1990}
\bibinfo{author}{\bibfnamefont{S.}~\bibnamefont{Sachdev}} \bibnamefont{and}
  \bibinfo{author}{\bibfnamefont{R.~A.} \bibnamefont{Jalabert}},
  \bibinfo{journal}{Mod. Phys. Lett. B} \textbf{\bibinfo{volume}{04}},
  \bibinfo{pages}{1043} (\bibinfo{year}{1990}).

\bibitem[{\citenamefont{Todo et~al.}(unpublished)\citenamefont{Todo, Matsuo,
  and Shitara}}]{TodoMS2012}
\bibinfo{author}{\bibfnamefont{S.}~\bibnamefont{Todo}},
  \bibinfo{author}{\bibfnamefont{H.}~\bibnamefont{Matsuo}}, \bibnamefont{and}
  \bibinfo{author}{\bibfnamefont{H.}~\bibnamefont{Shitara}}
  (\bibinfo{year}{unpublished}).

\bibitem[{\citenamefont{Bauer et~al.}(2011)\citenamefont{Bauer, Carr, Evertz,
  Feiguin, Freire, Fuchs, Gamper, Gukelberger, Gull, Guertler et~al.}}]{ALPS}
\bibinfo{author}{\bibfnamefont{B.}~\bibnamefont{Bauer}},
  \bibinfo{author}{\bibfnamefont{L.~D.} \bibnamefont{Carr}},
  \bibinfo{author}{\bibfnamefont{H.~G.} \bibnamefont{Evertz}},
  \bibinfo{author}{\bibfnamefont{A.}~\bibnamefont{Feiguin}},
  \bibinfo{author}{\bibfnamefont{J.}~\bibnamefont{Freire}},
  \bibinfo{author}{\bibfnamefont{S.}~\bibnamefont{Fuchs}},
  \bibinfo{author}{\bibfnamefont{L.}~\bibnamefont{Gamper}},
  \bibinfo{author}{\bibfnamefont{J.}~\bibnamefont{Gukelberger}},
  \bibinfo{author}{\bibfnamefont{E.}~\bibnamefont{Gull}},
  \bibinfo{author}{\bibfnamefont{S.}~\bibnamefont{Guertler}},
  \bibnamefont{et~al.}, \bibinfo{journal}{J. Stat. Mech.} p.
  \bibinfo{pages}{P05001} (\bibinfo{year}{2011}).

\bibitem[{\citenamefont{Harada}(2011)}]{HaradaKBayes2011}
\bibinfo{author}{\bibfnamefont{K.}~\bibnamefont{Harada}},
  \bibinfo{journal}{Phys. Rev. E} \textbf{\bibinfo{volume}{84}},
  \bibinfo{pages}{056704} (\bibinfo{year}{2011}).

\bibitem[{\citenamefont{Luijten et~al.}(1997)\citenamefont{Luijten, Bl\"ote,
  and Binder}}]{Luijten1997}
\bibinfo{author}{\bibfnamefont{E.}~\bibnamefont{Luijten}},
  \bibinfo{author}{\bibfnamefont{H.~W.~J.} \bibnamefont{Bl\"ote}},
  \bibnamefont{and} \bibinfo{author}{\bibfnamefont{K.}~\bibnamefont{Binder}},
  \bibinfo{journal}{Phys. Rev. E} \textbf{\bibinfo{volume}{56}},
  \bibinfo{pages}{6540} (\bibinfo{year}{1997}).

\bibitem[{\citenamefont{Wu}(1982{\natexlab{b}})}]{WuFY3D3Potts}
\bibinfo{author}{\bibfnamefont{F.~Y.} \bibnamefont{Wu}}, \bibinfo{journal}{Rev.
  Mod. Phys.} \textbf{\bibinfo{volume}{54}}, \bibinfo{pages}{235}
  (\bibinfo{year}{1982}{\natexlab{b}}).

\bibitem[{\citenamefont{Wu}(1983)}]{WuFY3D3Potts2}
\bibinfo{author}{\bibfnamefont{F.~Y.} \bibnamefont{Wu}}, \bibinfo{journal}{Rev.
  Mod. Phys.} \textbf{\bibinfo{volume}{55}}, \bibinfo{pages}{315(E)}
  (\bibinfo{year}{1983}).

\bibitem[{\citenamefont{Bl\"ote and Swendsen}(1979)}]{BloteHWJ1979}
\bibinfo{author}{\bibfnamefont{H.~W.~J.} \bibnamefont{Bl\"ote}}
  \bibnamefont{and} \bibinfo{author}{\bibfnamefont{R.~H.}
  \bibnamefont{Swendsen}}, \bibinfo{journal}{Phys. Rev. Lett.}
  \textbf{\bibinfo{volume}{43}}, \bibinfo{pages}{799} (\bibinfo{year}{1979}).

\bibitem[{\citenamefont{Janke and Villanova}(1997)}]{JankeW1997}
\bibinfo{author}{\bibfnamefont{W.}~\bibnamefont{Janke}} \bibnamefont{and}
  \bibinfo{author}{\bibfnamefont{P.}~\bibnamefont{Villanova}},
  \bibinfo{journal}{Nucl. Phys. B} \textbf{\bibinfo{volume}{489}},
  \bibinfo{pages}{679} (\bibinfo{year}{1997}).

\end{thebibliography}

\end{document}